\documentclass{aastex}
\usepackage{emulateapj5}

\def\kms{\relax \ifmmode {\,\rm km\,s}^{-1}\else \,km\,s$^{-1}$\fi}
 
\def\degree{\mbox{$^{\circ}$}}
\def\Mso{{M$_{\rm \odot}$}}
\def\mlr{\rm M$_\odot$~yr$^{-1}$}
\def\cm3{${\rm cm}^{-3}$}

\lefthead{Ram pressure stripping in Planetary Nebulae}
\righthead{Villaver, Garc\'{\i}a-Segura \& Manchado} 

\begin{document}

\title{Ram pressure stripping in Planetary Nebulae}
 
\author{Eva Villaver\altaffilmark{1,2}, Guillermo
  Garc\'{\i}a-Segura\altaffilmark{3}, \& Arturo Manchado\altaffilmark{1,4}}

\altaffiltext{1}{Instituto de Astrof\'{\i}sica de Canarias, V\'{\i}a
        L\'actea S/N, E-38200 La Laguna, Tenerife, Spain} 

\altaffiltext{2}{Space Telescope Science Institute, 3700 San Martin Drive,
  Baltimore, MD 21218; {\tt villaver@stsci.edu}}  

\altaffiltext{3}{Instituto de Astronom\'{\i}a-UNAM, Apartado postal 877,
       Ensenada, 22800 Baja California, M\'exico
 {\tt ggs@astrosen.unam.mx}}

\altaffiltext{4}{Consejo Superior de Investigaciones Cient\'{\i}ficas,
  Spain. {\tt amt@ll.iac.es}} 

\begin{abstract}
We present two-dimensional numerical simulations of the evolution of a
low-mass star moving supersonically through its surrounding interstellar
medium (ISM). We show that the ejecta of a moving star with a systemic
velocity of 20 \kms~will interact with the ISM and will form bow-shock 
structures qualitatively similar to what is observed.
We find that, due to ram-pressure 
stripping, most of the mass ejected during the AGB phase is left downstream
of the moving star. As a
consequence, the formation of the PN is highly influenced, even at the low
relative velocity of the star. The models are based on the predictions
of stellar evolution calculations. Therefore, the density and
velocity of the  
AGB and post-AGB winds are time dependent and give rise to the formation of
shock regions inside the cavity formed by the previous winds. As a result, the
stand-off distance is also time dependent and cannot be determined by simple
analytical arguments. 

\end{abstract}


\section{INTRODUCTION}

The process or processes that may cause the departure of spherical symmetry
in PNs have been the subject of many studies e.g. the hydrodynamic
collimation of the fast wind by an equatorial density enhancement
\citep{Fm:94}, the rotation of a single star with a magnetized wind
\citep{Rf:96,GGS:97,Glrf:99} 
and the evolution of a PN in a close binary system
\citep{Sok:96}. These mechanisms are able to account for the formation of
elliptical, point-symmetric and bipolar PNs. However, several PNs have been
observed in which mainly the external shell shows departure from
sphericity (e.g. MA 3, PM 1-295, M 2-40). In these PNs, the presence of
asymmetries cannot be explained by the physical processes related to the
central star (CS), such as rotation, magnetic fields or binarity. Instead,
the interaction with the ISM was proposed to be the cause of the observed
asymmetries.

The interaction of a PN with the ISM was first suggested by
\cite{Gur:69} and the first theoretical study was made by \cite{Smi:76}; 
he assumed a thin shell approximation and the `snow-plough' 
momentum-conserving model of \cite{Oor:51}. 
Later on, \cite{Isa:79}
used the same approximation, but considering higher velocities and ISM
densities. Both theoretical works arrived at a similar
conclusion, namely that the nebula fades away before any disruption of the
nebular shell becomes noticeable. \cite{Bss:90} analytically, and
\cite{Sbs:91} 
numerically confirmed the thin-shell approximation originally developed by
\cite{Smi:76}, concluding that the interaction becomes dominant when the
density of the nebular shell drops below a critical limit, due to the nebular
expansion. As a consequence, the interaction should be easily observable
during the late stages of the nebular evolution. Moreover, high ISM
densities, large relative velocities or both were needed to explain PN
asymmetries due to interaction with the ISM.

This is the scenario commonly accepted nowadays and as a result, the search
for interaction of PNs with the ISM has been mostly restricted to old PNs with
large angular extent \citep{Tk:96, Xetal:96}. Evidence for the interaction has
been also reported in the study of some individual objects (e.g.
\citealt{Bth:93, Tmn:95, Sz:97}) and the observations have been
interpreted in the light of the 
available models. As a consequence high spatial velocities were invoked to
explain the morphologies of PN at large distances from the Galactic plane.

The problem is that in the available models, the PN-ISM interaction 
was studied by considering the relative movement when nebular shell was
already formed. In this letter, we demonstrate that when the PN formation is 
followed by considering the evolution of the stellar wind from the AGB, high
velocities, high ISM densities or the presence of a 
magnetic field are not necessary to explain the observed asymmetries in the
external shells of PNs. Moreover, in contrast to what is commonly believed,
the interaction is observable even during the very early stages of the PN
phase. 

\section{THE NUMERICAL METHOD}

We have carried out the numerical simulations with the fluid
solver ZEUS-3D (\citealt{Sn:92a,Sn:92b,Smn:92}), developed by
M.~L. Norman and the Laboratory for 
Computational Astrophysics. The computations have been performed using a
2D spherical polar grid with the angular coordinate ranging from 0 to
180\degree~and a physical radial extension of 2~{\rm pc} which gives a total
grid size of 4$\times$2~{\rm pc}. We have adopted a resolution of
400$\times$360 zones in the radial and angular coordinates of the grid
respectively. We consider the photoionization of the gas during the PN phase
by using a simple approximation to derive the
location of the ionization front for arbitrary density distributions 
(see \citealt{Bty:79, Fttb:90}). This is done by assuming that ionization
equilibrium holds at all times and that the gas is fully ionized inside the H
II region. The models include the \cite{Rs:77} cooling curve above
10$^4$~{\rm K}. For temperatures below 10$^4$~{\rm K}, the unperturbed gas is
treated 
adiabatically but the shocked gas region is allowed to cool down with the 
radiative cooling curves given by \cite{Dm:72} and \cite{Mb:81}. The
photoionized gas is always kept at 10$^4$~{\rm K}, so no cooling curve is
applied to the H II region unless there is a shock. 

\subsection{The Boundary Conditions}
We assume that the ISM moves relative to the star perpendicular
to the line of sight by fixing the position of the star at
the center of the grid and allowing the ISM to flow into the grid at the outer
boundary from 0 to 90\degree. On the other hand,
from 90 to 180\degree~we set an outflow boundary condition. The temporal
evolution 
of the stellar wind for a 1 \Mso~star during the AGB phase (taken from
\citealt{Vw:93}) and the PN stage \citep{Vw:94} has been set within a small
(five radial zones) spherical region centered on the
symmetry axis, where reflecting boundary 
conditions are used. The gas evolution in a static configuration as well as
the stellar wind inputs for the models are described in detail in
\cite{Vgm:02} and \cite{Vmg:02}.   

In order to study the interaction process we have adopted a purely
hydrodynamical scheme, in which we assume that the ISM pressure is simply the
standard gas thermal pressure, $P = nKT$ (where {\rm n} is the number of
particles per unit volume, {\rm K} is the Boltzman's constant and {\rm T} is
the gas temperature). Since the strength of the interaction depends linearly
on the ISM density and quadratically on the relative velocity of the star, we
have adopted two ISM densities and a very conservative value for the star's
velocity, 20 \kms. The ISM is assumed to be homogeneous with the
characteristics of the  
warm neutral medium (WNM) which is the main observed constituent of the
ISM, with temperatures estimated to be in the range 5000-8000~{\rm K}
\citep{Kh:88} and neutral atomic hydrogen densities between 0.1 and 1
\cm3~\citep{Bur:88}. We use an ISM temperature of 6000~{\rm K} and
densities of 0.1 and 0.01 \cm3, the latter to study the interaction with a
very low density ISM. 

\section{RESULTS}

\subsection{The Evolution during the AGB through a Low Density Medium}

The logarithm of the gas density during the AGB phase is shown in
Figure~1 for a star moving with 20 \kms~through a low density ISM (n$_{\rm
o}$~=~0.01 \cm3). Figure~1 shows only half of the $r-\theta$ plane, the star
is fixed in the grid and the ISM flows in from the top to the bottom.  

During the very early stages of the evolution of an AGB star a free-streaming
steady stellar 
wind with a velocity of 2 \kms~and mass-loss rate of \.M = 10$^{-8}$
\mlr~reaches ram pressure equilibrium with the ISM at a stand-off distance
from the star of 0.13 {\rm pc} (first panel on the left of Fig.~1). After
this, the mass-loss rate and wind velocity are changing continuously in the
inner boundary giving rise to the formation of shocks 
that develop inside the bow-shock cavity formed by the early stationary
wind. From then onwards the stand-off distance cannot be computed from a
simple ram pressure balance argument between a free streaming stellar wind
and the ISM. The ram pressure of the ISM is balanced by the 
ram pressure of the stellar wind inside the
bow-shock cavity, making it a time dependent problem.

As a consequence of the interaction process a highly asymmetric shell is
formed that grows in size mainly in the downstream direction. 
The outer bow shock is radiative
for the stellar velocity considered, so the density of the swept-up shell is
proportional to the square of the upstream Mach number and
proportional to the ISM density. As the mass-loss rate continues (highly
modulated by the thermal pulses in the star, see Fig.~1 of Villaver et
al. 2002a) new shells are subsequently formed inside the structure generated
by the previous stellar wind. During most of the AGB evolution (355 000 {\rm
  yr} of the $\sim$ 495 000 {\rm yr} that the AGB phase lasts) the leading
part of the gas is not able to 
expand far away from the CS. Only the high mass-loss rate periods associated
with the last two thermal pulses during the AGB have
enough momentum to compete efficiently with the previously developed density
structure. It is at this moment that the leading parts of the outer shell
growin size. The last two panels show the density at the end of the second
thermal pulse and in the middle of the third one respectively. Because of the
relative movement, mass is constantly flowing away from the head of the bow
shock forming cometary structures behind the star.
 
\subsection{The Evolution during the AGB through a Higher Density Medium}
In Figure~2 we show the gas evolution for an ISM density of n$_o$~=~0.1
cm$^{-3}$, lower than the value in the Galactic plane ($\sim 1$ \cm3).
The snapshots of the evolution have been selected at the same times as those
shown in Figure~1. As expected, the 
effects of the stellar movement are more noticeable, since the ram pressure
exerted by the ISM depends linearly on the ISM density. The
compression and deformation of the upstream side of this shell is higher than
in the case shown in Figure~1. As for the low density case, the
stagnation radius can be computed from a ram pressure balance argument only
when the wind is stationary in the early evolution (see the first panel
of Fig.~2 where the stagnation radius is a factor 10$^{-1/2}$ smaller
than in the first panel of Fig.~1). 
In the present case, the upstream 
side of the shell does not increase in size until very late in the evolution
and the size is always at least 1.5 times smaller than the one formed when
the star evolves in a lower density medium. 

\subsection{The PN stage}
During the PN phase the stellar wind increases its velocity, generating an
adiabatic shock that forms the brightest shell of the nebula. In principle,
this shell should not be affected by the interaction with the ISM, unless
other mechanisms such as ambipolar diffusion under the presence of an external
magnetic field are invoked to propagate the asymmetry
to the innermost regions of the nebula. 

In the right panel of Figure~3 we show the gas density when the PN is
2~500~{\rm yr} old\footnote{That is 2~500~{\rm yr} after the photoionization
has started at the end of the AGB phase.} and when the ISM has a density
of 0.1~cm$^{-3}$. The simulated region has been reflected around the symmetry
axis in order to more clearly show the structure. The left panel shows the PN
at the same age, but when the star is 
evolving at rest with respect to the ISM (see Villaver et
al. 2002b). Two bright shells are clearly distinguished in both panels of
Figure 3. These are a spherical innermost shell shaped by the fast wind and
an external shell which shows a characteristic bow-shock configuration
in the right panel Figure~3.  

The first effect of the interaction, apart from that on the
morphology, is that the total size of the outer PN shell (halo) 
is reduced considerably. The radius of the halo in the left panel of Figure~3
is 1.8~{\rm pc}, whereas the outer shell of the interacting nebula has a
radius of 0.7~{\rm pc}. The displacement of the CS with respect to the
geometrical center of the nebula is 0.1 {\rm pc}, which corresponds to 20
\arcsec~at a 
distance of 1~{\rm kpc}. A less obvious effect is that the formation of the
outer shell of the PN takes place in a lower density environment, since most
of the mass ejected during the AGB phase is stripped by the ram pressure of
the ISM and left in the downstream direction of the stellar movement. The
model at rest has 0.7 \Mso~of gas above the ISM value in the 
circumstellar envelope (0.43~\Mso~that were lost by the star and 0.27 \Mso~of
ISM matter swept up by the wind), only 0.24~\Mso~of gas are recovered when the
star is moving. 

\section{DISCUSSION}

We find that the movement of the CS with respect to its surrounding medium
considerably alters the circumstellar gas structure formed during the AGB
phase and hence the PN formation. Up to now, it has been generally accepted
that the deceleration of the nebular shell due to its interaction with the
ISM can be more likely observed during the late stages of the nebular
evolution. We find that when the
evolution of the stellar wind is properly considered in the simulations, the
interaction with the ISM cannot be studied using simple ram pressure
balance arguments and plays a major role even during the early AGB 
evolution. 

From an observational point of view, the
presence of asymmetries in the haloes of PN 
that could be related to the interaction with the ISM is a common
feature\footnote{Multiple shells 
appear in 24\% of the spherical and
elliptical PN sample of the northern hemisphere \citep{Metal:96}; 40\% of
those show asymmetries in the haloes that could be related to the interaction
with the ISM.}. The high rate of observed asymmetries cannot be explained if
all of these PNs are moving at high velocities through high density
media. Instead, it can be understood when the evolution of the stellar
wind is taken into account, producing a much higher efficiency in the
interaction. Moreover, in agreement with our results,  
\cite{Gvm:98} show kinematical evidences that the interaction of
the external shells of PN with the ISM are found at all evolutionary stages.

According to our simulations it may not be necessary to invoke the presence of
a magnetic field in the ISM to explain the asymmetries found in the slow
moving (10-20 \kms) PNs Sh~2-216 \citep{Tmn:95} and NGC~6894 \citep{Zs:97}.
For the model parameters assumed in this paper, no instabilities appear in
the nebula, which is in agreement with the analytical results of 
\cite{Ds:94, Ds:98}. 

Another important aspect is that the interaction with the ISM
considerably reduces the mass of the circumstellar envelope during the AGB
and PN phases due to ram pressure stripping. Stellar evolution 
calculations predict that stars with initial masses 
in the range of $\sim$1--5~M$_\odot$ will end as PN nuclei with masses
around 0.6~M$_\odot$. Most of the mass is lost on the AGB
phase and should be easily observable as ionized mass during the PN
stage. However, observations of Galactic PNs reveal on average only 0.2
M$_\odot$ of ionized gas. The mass stripped away by the ISM when the star is
moving could solve the problem of the missing mass in PNs.

\section{SUMMARY}

We have studied the effects the relative movement of the CS with respect to
the ISM has on the evolution of the stellar ejecta during the AGB phase, and
on shaping the PN. Under the very conservative conditions assumed for the ISM
density and  the CS's velocity (20 \kms) we find that the PN-ISM interaction 
provides an adequate mechanism to explain the high rate of observed
asymmetries in the external shells of PNs. According to our simulations,
spherical haloes in PNs are expected only if the star is at rest with
respect to the ISM or if the star is moving at low angles with respect to the
line of sight, otherwise, bow-shock like and cometary structures should be
present. 
 
\acknowledgments 
We are grateful to the referee for very constructive suggestions. We thank
M. L. Norman and the Laboratory for Computational Astrophysics for the use of
ZEUS-3D. EV would like to thank Tariq Shahbaz and M\'onica S\'anchez Cuberes
for fruitful discussions. GGS 
is partially supported by DGAPA-UNAM grant IN117799 and CONACyT.

\newpage
\begin{figure}
\epsscale{0.8}
\plotone{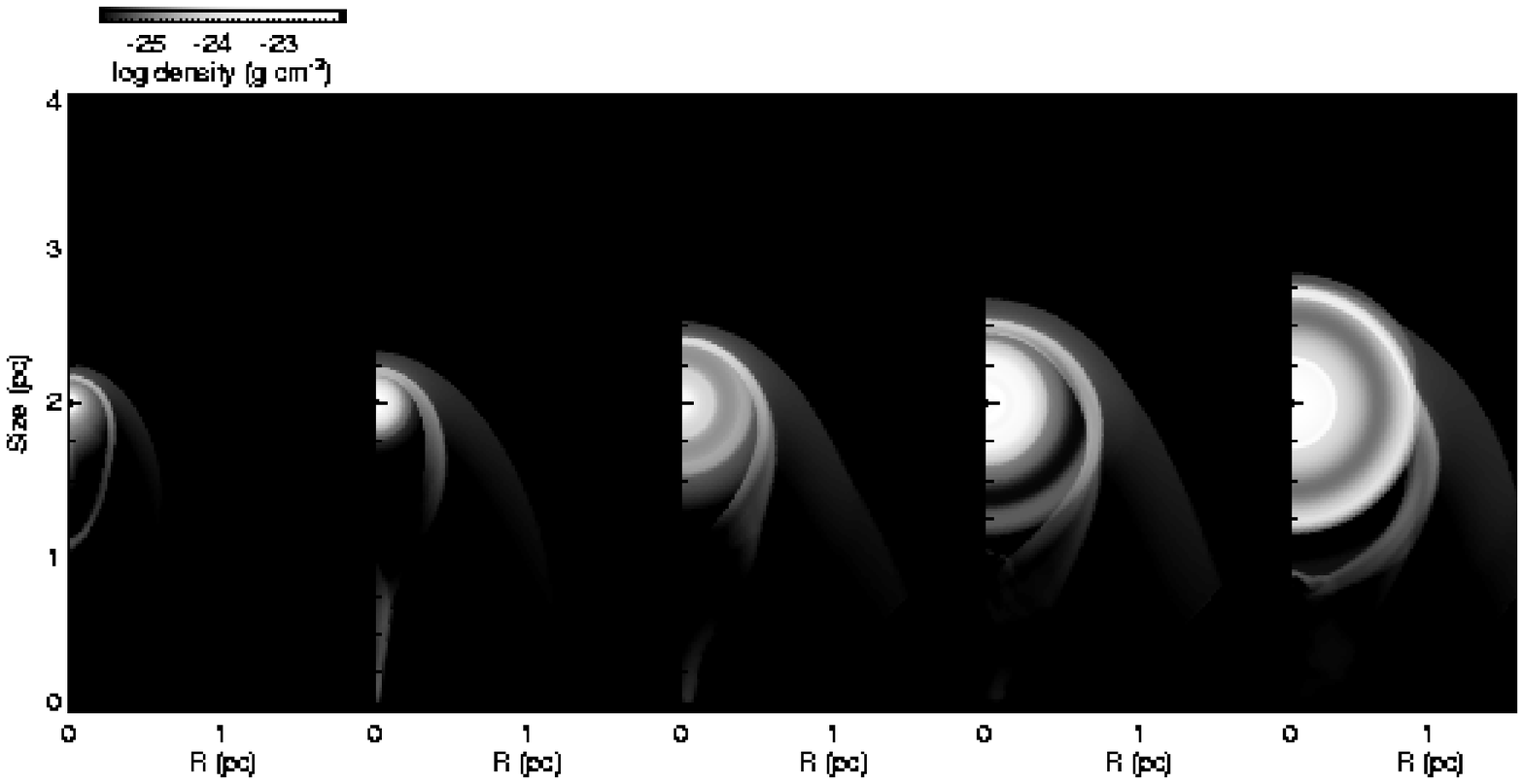}
\caption{The gas density of the structure
generated by a star during the AGB phase moving with a velocity of 20
\kms~through an ISM 
with a density of n$_{\rm o}$=0.01 cm$^{-3}$. The intensity scale is
logarithmic.  
All the densities below and above
the values shown in the color scale have been saturated. The first panel on
the left corresponds to 85 000 {\rm yr} and the subsequent panels are shown
at 206~000, 327~000, 387~400 and 447~800 {\rm yr} respectively, all of them
during the AGB phase.}
\end{figure}

\begin{figure}
\epsscale{0.8}
\plotone{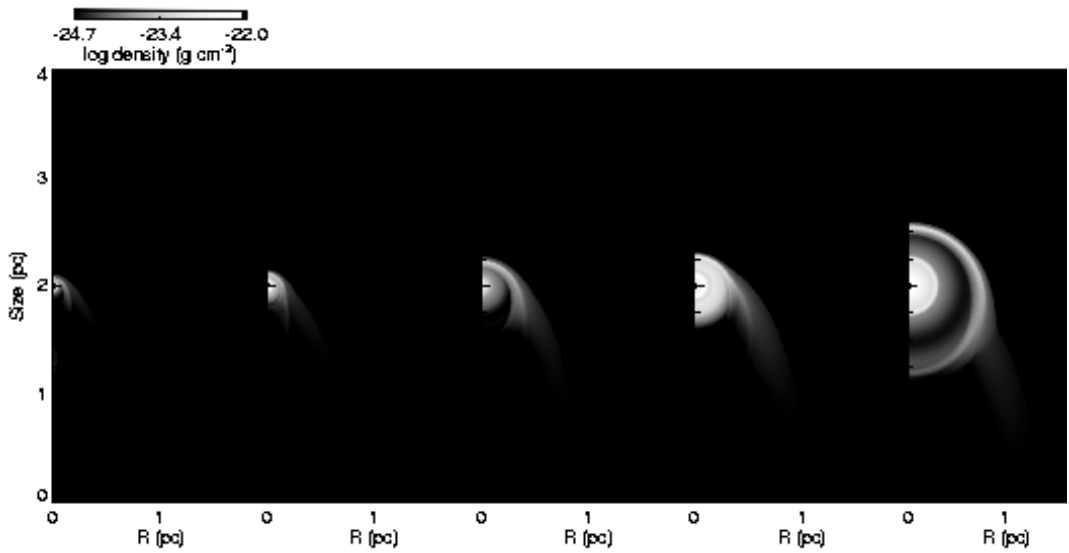}
\caption{Same as Figure 1, but for an ISM with a density of 0.1 \cm3.}
\end{figure}

\begin{figure}
\epsscale{0.7}
\plotone{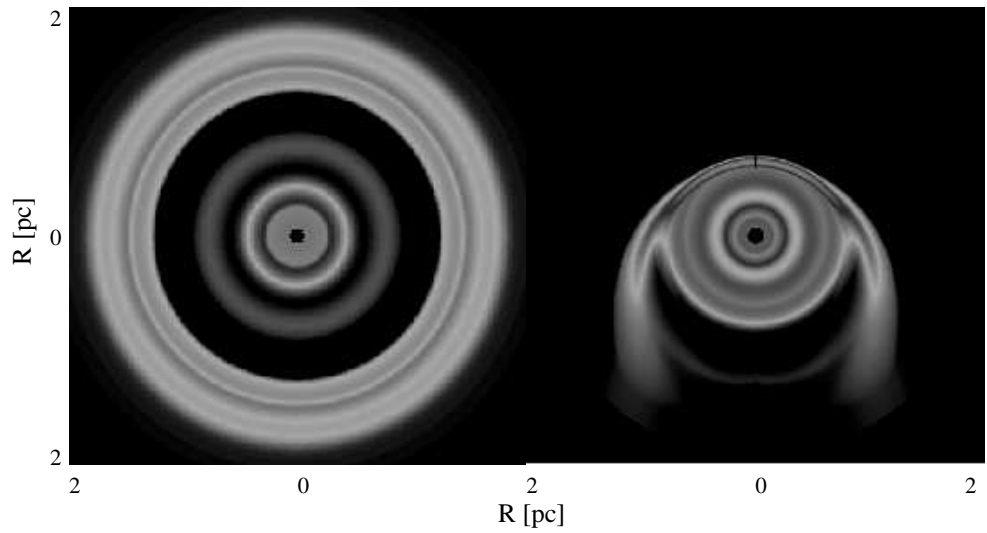}
\caption[ ]{The gas density of a PN 2500 {\rm yr} old when the 
the star is at rest (left panel) and 
when the star is moving at 20 \kms~through an ISM
with a density of 0.1 \cm3~(right panel).
The intensity scale is logarithmic.
\label{fig3.eps}}
\end{figure}


\begin{thebibliography}{}
\bibitem[Bodenheimer, Tenorio-Tagle, \& Yorke(1979)]{Bty:79}
Bodenheimer, G., Tenorio-Tagle, G., \& Yorke, H. W. 1979, \apj, 233, 85
\bibitem[Borkowski, Sarazin, \& Soker(1990)]{Bss:90}
Borkowski, K. J., Sarazin, C. L., \& Soker, N. 1990, \apj, 360, 173
\bibitem[Borkowski, Tsvetanov, \& Harrington(1993)]{Bth:93}
Borkowski, K. J., Tsvetanov, Z. \& Harrington, J. P. 1993, \apj, 402, 57
\bibitem[Burton(1988)]{Bur:88}
Burton, W. B. 1988.  In Galactic and Extragalactic Radio
Astronomy, ed. K. Kellermann, G. L. Verschuur (New York: Springer), 295

\bibitem[Dalgarno \& McCray(1972)]{Dm:72}
Dalgarno, A. \& McCray, R.A. 1972, ARA\&A, 10, 375

\bibitem[Dgani \& Soker(1994)]{Ds:94}
Dgani, R. \& Soker, N. 1994, \apj, 434, 262
\bibitem[Dgani \& Soker(1998)]{Ds:98}
Dgani, R. \& Soker, N. 1998, \apj, 495, 337

\bibitem[Franco, Tenorio-Tagle, \& Bodenheimer(1990)]{Fttb:90}
Franco, J., Tenorio-Tagle, G., \& Bodenheimer, P. 1990, \apj, 349, 126

\bibitem[Frank, \& Mellema(1994)]{Fm:94}
Frank, A., \& Mellema, G. 1994, \apj, 430, 800

\bibitem[Garc\'{\i}a-Segura et al.(1999)]{Glrf:99}
Garc\'{\i}a-Segura, G., Langer, N., R\'zyczka, M., \& Franco, J. 1999, 
\apj, 517, 767

\bibitem[Garc\'{\i}a-Segura (1997)]{GGS:97}
Garc\'{\i}a-Segura, G. 1997, \apj, 489, L189 

\bibitem[Guerrero, Villaver, \& Manchado(1998)]{Gvm:98}
Guerrero, M. A., Villaver, E., \& Manchado, A. 1998, \apj, 507, 889
 
\bibitem[Gurzadyan(1969)]{Gur:69}
Gurzadyan, G. A. 1969, New York Gordon and Breach p.235


\bibitem[Isaacmann(1979)]{Isa:79}
Isaacmann R. 1979, A\&A, 77, 327


\bibitem[Kulkarni \& Heiles(1988)]{Kh:88}
Kulkarni, S.~R. \& Heiles, C. 1988, In Galactic and Extragalactic Radio
Astronomy, ed. K. Kellermann, G. L. Verschuur (New York: Springer), 95
\bibitem[MacDonald \& Bailey(1981)]{Mb:81}
MacDonald, J. \& Bailey, M. E. 1981, \mnras, 197, 995

\bibitem[Manchado et al.(1996)]{Metal:96}
Manchado, A., Guerrero, M. A., Stanghellini, L., \& Serra-Ricart, M. 1996, 
The IAC Morphological Catalog of Northern Galactic Planetary Nebulae 
(La Laguna: IAC) (MGSS)

\bibitem[Oort(1951)]{Oor:51}
Oort, J. H. 1951, {\it Problems of cosmical aerodynamics, central Air
  Document Office, Dayton.}

\bibitem[Raymond \& Smith(1977)]{Rs:77}
Raymond, J.~C. \& Smith, B.~W. 1977, \apjs, 35, 419

\bibitem[R\'o\.zyczka \& Franco(1996)]{Rf:96}
R\'o\.zyczka, M. \& Franco, J. 1996, ApJL, 469, L127

\bibitem[Smith(1976)]{Smi:76}
Smith, H. 1976, \mnras, 175, 419

\bibitem[Soker(1996)]{Sok:96}
Soker, N. 1996, \apj, 496, 734

\bibitem[Soker, Borkowski, \& Sarazin(1991)]{Sbs:91}
Soker, N., Borkowski, K.J., \& Sarazin, C.L. 1991, \aj, 102, 1381

\bibitem[Soker \& Zucker(1997)]{Sz:97}
Soker N. \& Zucker, D. B. 1997, \mnras, 289, 665


\bibitem[Stone \& Norman(1992a)]{Sn:92a}
Stone, J.~M. \& Norman, M. L. 1992a, \apjs, 80, 753
\bibitem[Stone \& Norman(1992b)]{Sn:92b}
Stone, J.~M. \& Norman, M. L. 1992b, \apjs, 80, 791
\bibitem[Stone, Mihalas, \& Norman(1992)]{Smn:92}
Stone, J.~M., Mihalas, D., \& Norman, M. L. 1992, \apjs, 80, 819

\bibitem[Tweedy, Martos, \& Noriega-Crespo(1995)]{Tmn:95}
Tweedy, R. W., Martos, M. A. \& Noriega-Crespo, A. 1995, \apj, 447, 257

\bibitem[Tweedy \& Kwitter(1996)]{Tk:96}
Tweedy, R. W. \& Kwitter, K. B. 1996, \apjs, 107, 255

\bibitem[Vassiliadis \& Wood(1993)]{Vw:93}
Vassiliadis, E. \& Wood, P. 1993, \apj, 413, 641

\bibitem[Vassiliadis \& Wood(1994)]{Vw:94}
Vassiliadis, E. \& Wood, P. 1994, \apj, 92, 125

\bibitem[Villaver, Garc\'{\i}a-Segura, \& Manchado(2002a)]{Vgm:02}
Villaver, E., Garc\'{\i}a-Segura, \& Manchado, A. 2002, \apj, 571, 880 

\bibitem[Villaver, Manchado \& Garc\'{\i}a-Segura(2002b)]{Vmg:02}
Villaver, E., Manchado, A., \& Garc\'{\i}a-Segura 2002, \apj, 581, 1204 

\bibitem[Xilouris et al.(1996)]{Xetal:96}
Xilouris, K. M., Papamastorakis, J., Paleologou, E.\& Terzian, Y. 1996, A\&A
603,612

\bibitem[Zucker \& Soker(1997)]{Zs:97}
Zucker, D. B., \& Soker, N. 1997, \mnras, 289, 665
\end{thebibliography}
\end{document}